\documentclass[prl,superscriptaddress,twocolumn]{revtex4}

\pdfoutput=1

\usepackage{amsmath}
\usepackage{amsfonts}
\usepackage{amssymb}
\usepackage[pdftex]{graphicx}
\usepackage[pdftex]{hyperref}
\usepackage{color}

\begin{document}

\title{Nonlinear terahertz metamaterials via field-enhanced carrier dynamics in GaAs}
\date{\today}\received{}
\author{Kebin Fan}
\affiliation{Department of Mechanical Engineering, Boston University, 110 Cummington Street, Boston, Massachusetts 02215, USA}
\author{Harold Y. Hwang} \affiliation{Department of Chemistry, Massachusetts Institute
of Technology, Cambridge, Massachusetts 02139, USA}
\author{Mengkun Liu}
\affiliation{Department of Physics, Boston University, 590
Commonwealth Avenue, Boston, Massachusetts, 02215, USA}
\author{Andrew C. Strikwerda}
\affiliation{Department of Physics, Boston University, 590
Commonwealth Avenue, Boston, Massachusetts, 02215, USA}
\author{Aaron Sternbach}
\affiliation{Department of Physics, Boston University, 590
Commonwealth Avenue, Boston, Massachusetts, 02215, USA}
\author{Jingdi Zhang}
\affiliation{Department of Physics, Boston University, 590 Commonwealth Avenue, Boston, Massachusetts, 02215, USA}
\author{Xiaoguang Zhao}
\affiliation{Department of Mechanical Engineering, Boston University, 110 Cummington Street, Boston, Massachusetts 02215, USA}
\author{Xin Zhang}
\affiliation{Department of Mechanical Engineering, Boston
University, 110 Cummington Street, Boston, Massachusetts 02215,
USA}
\author{Keith A. Nelson}
\affiliation{Department of Chemistry, Massachusetts Institute of Technology, Cambridge, Massachusetts 02139, USA}
\author{Richard~D.~Averitt}
\affiliation{Department of Physics, Boston University, 590 Commonwealth Avenue, Boston, Massachusetts, 02215, USA}

\begin{abstract}
We demonstrate nonlinear metamaterial split ring resonators (SRRs)
on GaAs at terahertz frequencies. For SRRs on doped GaAs films,
incident terahertz radiation with peak fields of $\sim$20 - 160 kV/cm drives
intervalley scattering. This reduces the carrier mobility and
enhances the SRR LC response due to a conductivity decrease in the
doped thin film. Above $\sim$160 kV/cm, electric field
enhancement within the SRR gaps leads to efficient impact
ionization, increasing the carrier density and the conductivity which, in turn, suppresses
the SRR resonance. We demonstrate an increase of up to 10 orders of magnitude
in the carrier density in the SRR gaps on semi-insulating GaAs substrate. Furthermore, we show that the effective
permittivity can be swept from negative to positive values with increasing
terahertz field strength in the impact ionization regime, enabling new
possibilities for nonlinear metamaterials.

\end{abstract}

\maketitle

Nonlinear metamaterials is a rapidly developing field of fundamental
interest with significant technological implications spanning from
microwave through the visible spectral ranges \cite{zharov03, klein07, bing08,
fang09, poutrina10, rose11a, liu12}. As with tunable and reconfigurable
metamaterials \cite{chen06, tao09}, the combination of the
metamaterial structure with the local environment is crucial. This
is because significant nonlinearities result from local field
enhancement within the active region of the subwavelength
metamaterial elements which, in the case of split ring resonators,
are the capacitive gaps. While the active volume of the enhanced
gaps is small in comparison to the unit cell volume, the field
enhancement can dominate volumetric effects leading to
global nonlinearities enhanced by two to four orders of magnitude
\cite{poutrina10}. This, in turn, results in useful nonlinear
effects at low incident fields.

Advances in nonlinear metamaterials coincide with the development of
high-field terahertz sources capable of generating electric fields
sufficient to induce significant nonlinearities in conventional matter
\cite{matthias09, hebling10, ic11, tanaka11, hirori11}. For example,
in doped GaAs, highly nonlinear effects such as velocity saturation
and impact ionization have been observed \cite{hebling10, hirori11}
at peak electric fields of several hundred kV/cm. Further, with metamaterial
THz field enhancement to MV/cm  fields an insulator-metal phase
transition has been induced in vanadium dioxide, a prototypical
 correlated electron material \cite{liu12}.

In this letter, we experimentally demonstrate a nonlinear response
in metamaterial split ring resonators (SRRs) on n-type GaAs and
semi-insulating (SI) GaAs at terahertz frequencies. The nonlinear
response arises from THz electric field-induced carrier dynamics
that increase or decrease the substrate conductivity upon which the
SRR arrays are fabricated. This modifies the SRR electromagnetic
response as a function of field strength. For peak incident THz
fields ($E_{in}$) from $\sim$20-160 kV/cm, mobility saturation by intervalley scattering
(IVS) dominates leading (for doped GaAs) to a conductivity decrease
and a corresponding increase in the metamaterial oscillator
strength. In this regime, electric field enhancement within the SRR
capacitive gaps does not significantly contribute to the
nonlinearity. That is, IVS occurs uniformly across the film
independent of the SRRs. For $E_{in}$ $>$ 160 kV/cm, field
enhancement in the gaps enables efficient carrier generation via impact ionization (IMI) resulting
in a decrease in the SRR oscillator strength. It is possible to sweep
the effective permittivity from negative to positive values by
increasing the incident field strength enabling new possibilities for nonlinear
metamaterials such as an intensity dependent
effective refractive index. We also show
that SRRs on semi-insulating GaAs exhibit a nonlinear response arising from an
increase of the in-gap conductivity by several orders of magnitude.

\begin{figure}[b]
\begin{center}
\includegraphics[width=3.5in,keepaspectratio=true]%
{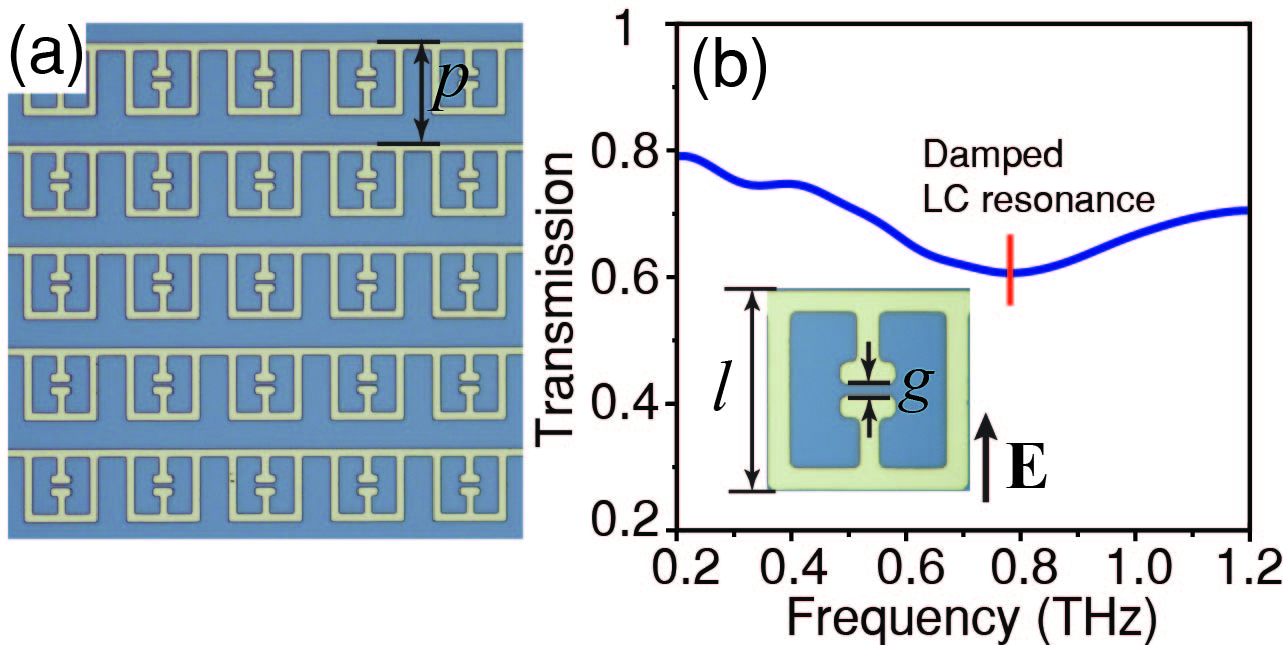}%
\caption{(a). Image of the SRR array on n-type GaAs substrate. The
period $p$ is 50 $\mu m$; the length of ring $l$ is 36 $\mu m$; the
gap $g$ is 2.2 $\mu m$. (b) Field transmission spectrum of SRR array using
THz time domain spectroscopy on doped GaAs
at low fields (i.e. in the linear response regime).}%
\label{Figure1}%
\end{center}
\end{figure}

To create nonlinear metamaterials, electric field resonant SRR
arrays were fabricated by conventional microfabrication on GaAs.
Fig. 1 (a) shows an SRR array on a 1.8-$\mu$m-thick doped
($1\times10^{16}\ cm^{-3}$) GaAs film deposited on a SI-GaAs
substrate using molecular-beam epitaxy. The SRRs consist of 200 nm
thick Au with a 10 nm Cr adhesion layer. A narrow gap ($g$=2.2 $\mu$m)
is important as the field enhancement
scales as 1/$g$ \cite{werley12}. To characterize the linear
electromagnetic (EM) response, electro-optic based terahertz
time-domian (THz-TDS) spectroscopy was employed. To accurately
measure the metamaterial transmission spectrum, a bare n-type GaAs substrate
was used as a reference. Dividing the Fourier transform of the
sample data by that of the reference data yields the transmission amplitude as a
function of frequency as shown in Figure \ref{Figure1}(b). The LC
resonance is $\sim$ 0.75 THz and is strongly damped, as expected,
since carriers in the doped GaAs short the SRRs. The corresponding
conductivity of doped GaAs is 7 $(\Omega cm)^{-1}$.

For the nonlinear transmission measurements, high-field THz
experiments were performed using tilted-pulse-front THz generation
in LiNbO$_{3}$ \cite{yeh07}. We obtained a peak THz field strength of approximately
400 kV/cm, with a 1 mm beam diameter at the focus. The field
strength could be varied from 24 kV/cm to 400 kV/cm using a pair of
wire grid polarizers. The detailed experimental setup has been
described elsewhere \cite{hebling08, matthias09}. All experiments
were performed at room temperature in a dry air environment at
ambient pressure.

%\begin{figure*}[htbp]
%\begin{center}
%\includegraphics[width=3.5in,keepaspectratio=true]%
\begin{figure}[b]
\begin{center}
\includegraphics[width=3.5in,keepaspectratio=true]%
{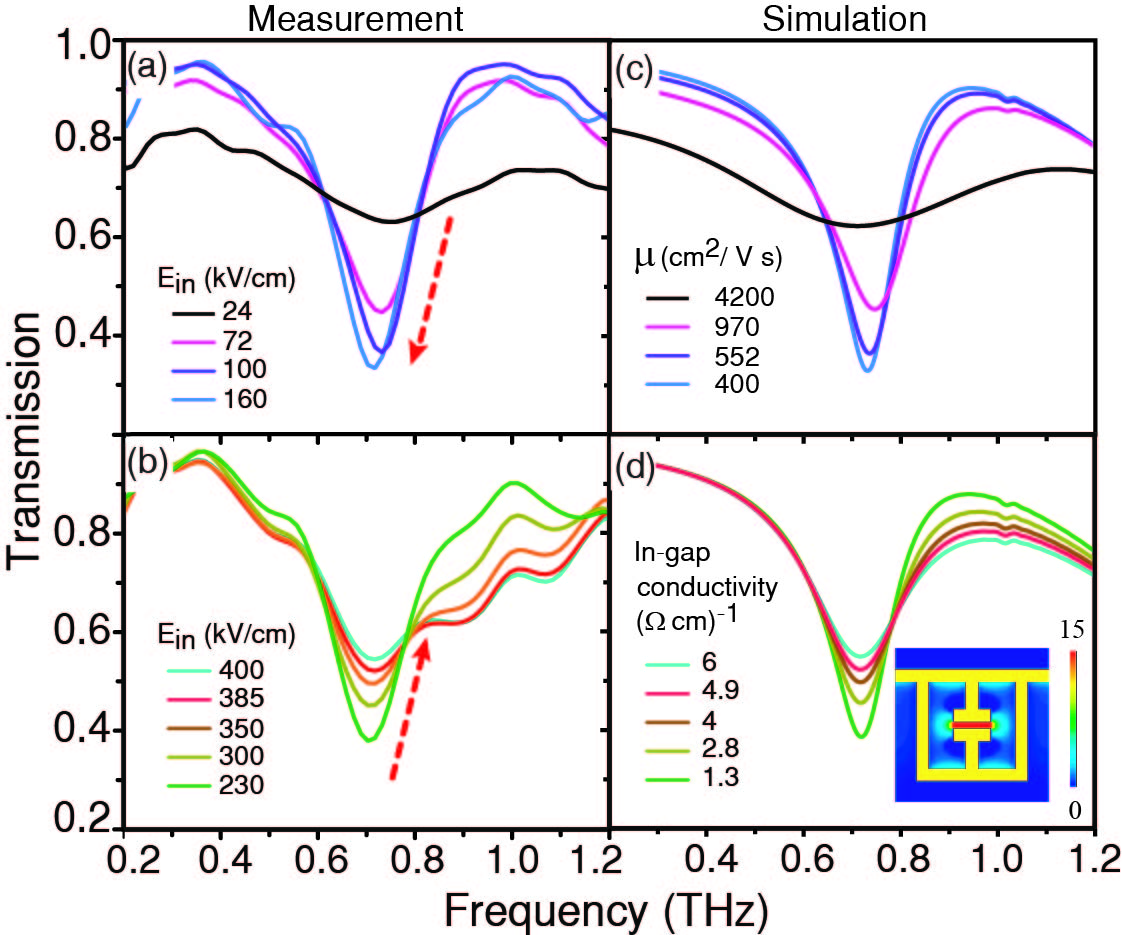}%
\caption{(a) \& (b) Experimental electric field transmission as a
function of frequency for various incident peak electric fields. (c)
Simulated transmission of the SRR array on doped GaAs
film for different electron mobilities at constant carrier density of
$1\times10^{16} cm^{-3}$. In these simulations the mobility of the carriers
throughout the film is changed. $\mu$ is the
electron mobility. (d) Simulated transmission on doped
GaAs for various in-gap conductivities. In these simulations the mobility is constant and the
carrier density is modified according to IMI simulations. The inset in (d) shows the
simulated on-resonance field enhancement highlighting that the largest enhancement
occurs within the SRR gaps.}%
\label{Figure2}%
\end{center}
\end{figure}
%\end{figure*}

Figure \ref{Figure2}(a) shows the experimentally measured
transmission as a function of frequency for various values of
$E_{in} \leq$ 160 kV/cm. At 24 kV/cm the transmission response is similar to the low-field ($<$1 kV/cm)
result shown in Fig. \ref{Figure1}(b) where a weak LC resonance is
observed. As the incident field is increased to 160 kV/cm, the LC resonance becomes
quite pronounced leading to a decrease in the resonant
transmission($\thicksim$ 0.73 THz) from 65\% to about 30\%. This
suggests a decrease in the conductivity which, as discussed below,
results from a decrease in the carrier mobility due to THz-induced
IVS. The off-resonance transmission at low frequencies
increases (e.g. at 0.4 THz from 80\% to over 90\%). This indicates a
field-induced reduction in the conductivity of the entire n-type
GaAs film similar to other tunable metamaterial responses
\cite{driscoll08, shen11} and consistent with earlier observation
in bulk GaAs \cite{hebling10}. Thus, for 24 $\leq $ $E_{in}$ $\leq$
160 kV/cm the nonlinearity does not depend strongly on the in-gap
field enhancement.

However above $\thicksim$ 160 kV/cm, the trend described above in the field-dependent transmission
at the resonant frequency is exactly reversed as shown in Fig. 2(b):
the transmission decreases with further increase in $E_{in}$.
This indicates a reversal of the underlying trend in conductivity as
well, i.e. an increase in conductivity in the gaps as the field is
increased above 160 kV/cm. Notably, the off-resonance transmission at
lower frequencies ($<$ 0.5 THz) remains almost unchanged and implies
a local conductivity increase solely in the capacitive gaps of the resonators.
We ascribe this response to THz field-induced impact ionization.

To better understand how THz field-driven nonequilibrium carrier
transport determines the nonlinear metamaterial response, we
consider, in turn, IVS and IMI. For 24 $\leq$ $E_{in}$ $\leq$ 160
kV/cm, electrons in the conduction band ($\Gamma$ valley) are
accelerated and acquire energy on the order of 1 eV, sufficient for
scattering into satellite valleys. Since the mobility in the side
valleys is approximately one order of magnitude smaller than in the
$\Gamma$ valley \cite{darrow92, nuss87}, IVS leads to a significant conductivity decrease in the
n-type GaAs film. Independent of the SRR array, for doped
substrates IVS results in photoinduced transparency
\cite{hebling10,fhsu09,allenAPL83} and accounts for the off-resonance
transmission increase that is observed in Fig. 2(a). Of course, for
the SRR arrays, this same process decreases the in-gap conductivity,
resulting in an enhanced LC resonance.

IVS-induced enhancement of the metamaterial EM response was simulated
using a commercial software, CST Microwave Studio. The n-type GaAs film was modeled
utilizing a frequency dependent Drude model.
We assume that the average mobility varies by changing the
relative fraction of electrons in the $\Gamma$ and L valleys
and that at 160 kV/cm all of the conduction electrons
are scattered to the L valley. Additionally, for fields $\leq$ 160 kV/cm, we
neglect IMI. That is, the carrier concentration is
held constant at the doping value ($1\times10^{16}$ $cm^{-3}$). The
results of this mobility averaging procedure (applied to the entire
film) determine the film conductivity used in the electromagnetic
simulations. The results are shown in Fig. 2(c) and show good
agreement with the experimental data in Fig. 2(a). Comparing these
plots reveals that, for an incident field of 24 kV/cm, a mobility
of 4200 cm$^{2}/Vs$ (close to the reported value
\cite{katz92}) yields good agreement. If the majority of conduction
electrons were scattered to the L valley, a mobility of 400
cm$^{2}/Vs$ would result. The agreement between experiment and
simulation suggests that at 160 kV/cm this a reasonable
approximation.

Above 160 kV/cm, IMI becomes increasingly important.
This is because the SRR resonance is strengthened
from the IVS-induced conductivity decrease. At 160 kV/cm, simulations indicate in-gap
electric field enhancement by a factor of $\sim$15 on resonance, corresponding to
an overall peak electric field enhancement of $\sim$ 4 in the
time domain. This is in good agreement with previously reported values
\cite{werley12}. Thus, with an incident field strength of 160 kV/cm,
the in-gap field is estimated to be $\sim$0.64 MV/cm,
which is sufficient for IMI in which THz-accelerated conduction band electrons
collide with valance band electrons to result in carrier generation (in
the L valley) \cite{tanaka11, hirori11}. As a
simple estimate, the motion of electrons under an external field is
given as
\begin{eqnarray}
\frac{{d}{\langle v(t)\rangle}}{dt}=
 \frac{{e}{E(t)}}{{m}_{e}}-\frac{\langle v(t)\rangle}{\tau{(E)}}
 \label{stokes}
\end{eqnarray}

\noindent where $\langle v(t)\rangle$ is the average electron velocity, $E(t)$
is the THz field, ${m}_{e}$ is the effective electron mass, and
$\tau$ is the relaxation time \cite{kuehn}. For n-type
GaAs, IMI is favorable for electrons in the L valley when
the kinetic energy of an electron reaches a threshold energy,
$\mathcal{E}_{th}$ = 2.1 eV \cite{anderson72}. Model calculations show that
with an initial carrier density of $1 \times 10^{16}$ $/cm^3$, an
incident field of 400 kV/cm (1.6 MV/cm in gap field) can increase
the carrier density up to $\sim$ 10$^{17}$ $/cm^3$.
 This process leads to an in-gap conductivity of $\sim$ 6
$(\Omega cm)^{-1}$. The calculated conductivity arising from IMI
leads to reasonable agreement between experiment and
simulation (Fig. \ref{Figure2}(b) and (d)). For
agreement between experiment and simulation at the highest fields,
an average in-gap electron mobility of 240 cm$^{2}/Vs$ was used, consistent with increased carrier-carrier
scattering at higher densities and in  agreement with the lower end
of the L valley mobility previously reported in optical carrier
generation studies \cite{nuss87,darrow92}.

\begin{figure}[htb]
\begin{center}
\includegraphics[width=2.5in,keepaspectratio=true]%
{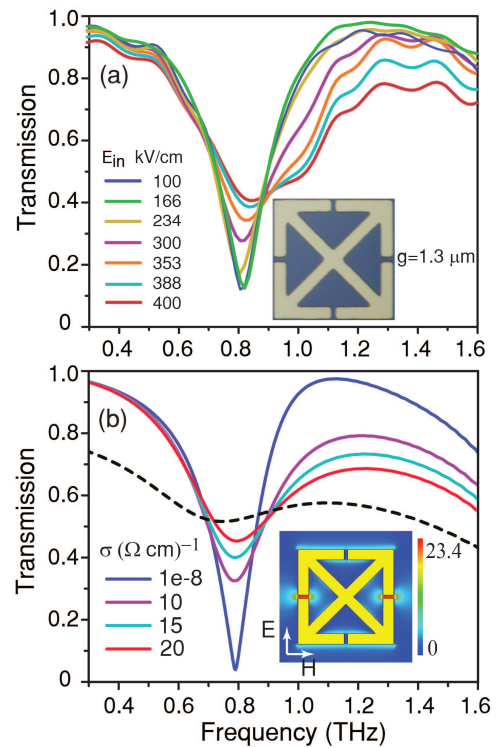}%
\caption{(a) Experimental electric field transmission as a function
of frequency for various incident peak electric fields for SRRs on
SI-GaAs. (b) Simulated transmission for SRRs on SI-GaAs as a
function of frequency for various in-gap conductivities. The dashed
curve is the predicted response if the entire lateral area of the GaAs substrate (to a depth of 0.5 $\mu$m)
has a conductivity of 20 $(\Omega cm)^{-1}$. This clearly disagrees with experiment suggesting that the
carrier multiplication occurs only in the capacitive gaps of the SRRs. The inset shows the simulated on-resonance field enhancement.}%
\label{Figure3}%
\end{center}
\end{figure}

In contrast to doped GaAs, semi-insulating GaAs (SI-GaAs, carrier density about
$1 \times 10^7$ $/cm^3$) with its low conductivity of $1 \times 10^{-8}$ $(\Omega cm)^{-1}$
yields a distinct SRR resonance and THz field enhancement even at low
incident field levels. The inset of Fig. \ref{Figure3} shows the
SRR unit cell used to fabricate arrays on SI-GaAs. The
size and period of the SRR resonators are the same as the resonators
on the doped GaAs, thereby maintaining the resonance roughly
at the same frequency. A smaller SRR gap (1.3 $\mu m$) was
used to increase the field enhancement given the difficulty in
initiating IMI on SI-GaAs due to the small initial carrier density.
At our free-space THz field strengths, IMI is ineffective as a carrier generation
mechanism in SI-GaAs. However, with the additional
field enhancement available from SRRs, we observed a nonlinear
response on SI-GaAs that we attribute mainly to IMI.

Figure \ref{Figure3}(a) shows the experimentally measured
transmission of metamaterials as a function of frequency for various
values of E$_{in}$. At the lowest field (100 kV/cm),
the response is still in the linear regime with a characteristic dip in
the transmission at the LC resonance frequency. As the incident
field increases, the metamaterial transmission increases and
broadens with negligible change at lower frequencies. We
note that even at the highest incident fields (E$_{in}$ = 400 kV/cm), no nonlinear
transmission changes were detected on bare SI-GaAs. Similar to the
nonlinear high field results ($>$ 160 kV/cm) on n-type GaAs, the
results of Fig. 3(a) suggest a large increase in the in-gap conductivity,
indicating a carrier density increase via IMI. EM simulations were performed by changing the
conductivity of the GaAs in the horizontal gaps as shown in the inset
of Fig. 3(b). The plot shows the transmission change upon increasing
the in-gap conductivity from $10^{-8}$ $(\Omega cm)^{-1}$ to
20 $(\Omega cm)^{-1}$. The simulations show a trend that is markedly
similar to the experimental results.

The inset of Fig. \ref{Figure3}(b) shows the field within the gaps
is enhanced by over 20 times on resonance corresponding to a peak time-domain field
enhancement factor of 7. Impact ionization calculations show that with an
initial carrier density of $1 \times 10^{7}$ $/cm^3$, an in-gap field
of 2.8 MV/cm (corresponding to our largest incident field, E$_{in}$ = 400 kV/cm)
can increase the carrier density by more than ten orders of magnitude up to $\sim$ 10$^{18}$ $/cm^3$.
Although we attribute the carrier multiplication primarily to IMI, Zener tunneling cannot, a priori, be discounted.
Tunneling calculations \cite{rodin02} show that with an in-gap peak field
of 2.8 MV/cm, Zener tunneling can generate a free carrier density of $\sim$ $10^{17}$ $/cm^3$ which
is one order of magnitude smaller than that reached by IMI. This indicates that carrier generation may
involve both processes.

\begin{figure}[htb]
\begin{center}
\includegraphics[width=3.4in,keepaspectratio=true]%
{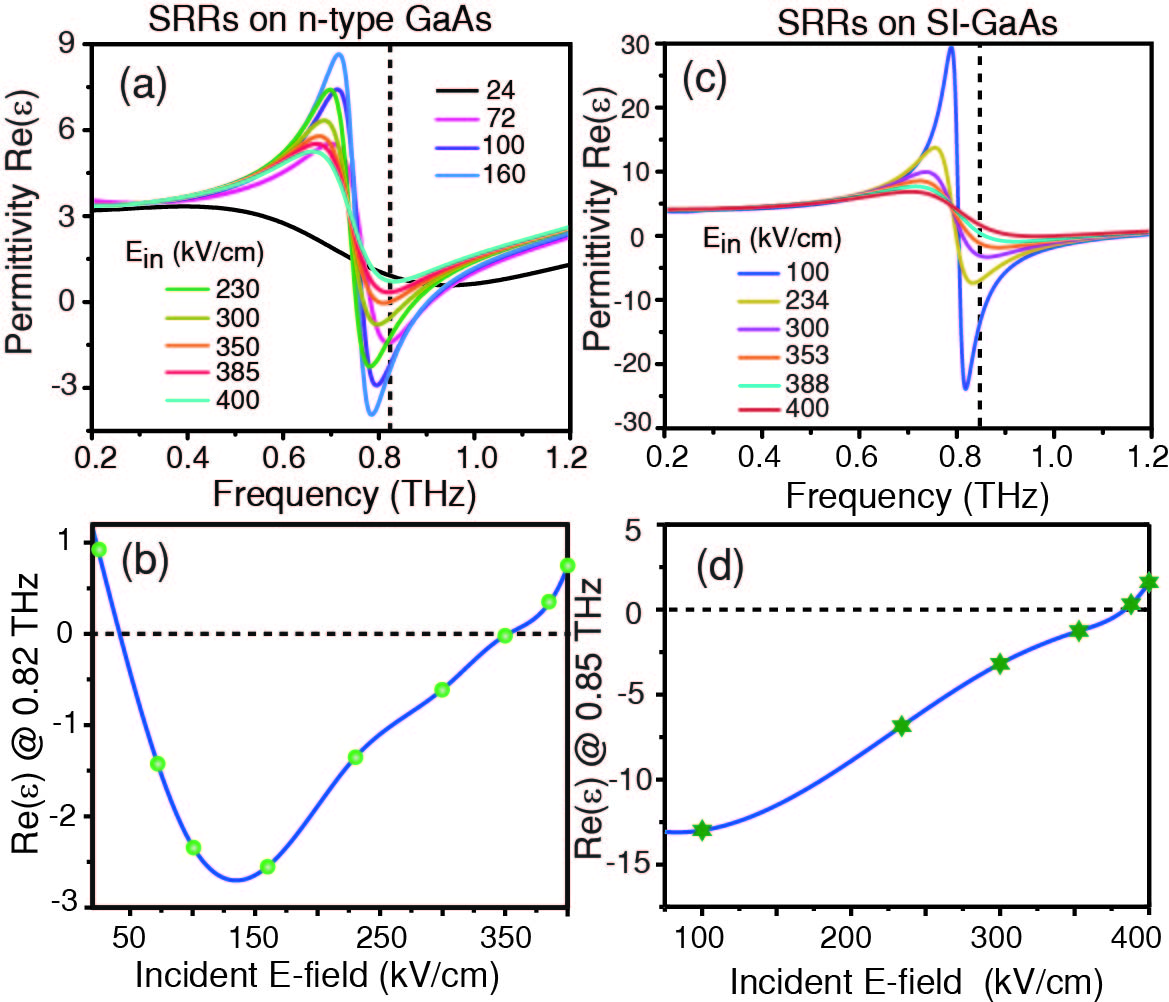}%
\caption{(a) Effective real permittivity $\epsilon_{re}(\omega)$
for SRRs on doped GaAs for various incident electric fields.
(b) $\epsilon_{re}(\omega)$ as a function of incident peak electric
field strength at 0.82 THz for SRRs on doped GaAs. (c)
$\epsilon_{re}(\omega)$ for SRRs on SI-GaAs for various incident
electric fields. (d) $\epsilon_{re}(\omega)$ as a function of
incident peak electric field strength at 0.85 THz for SRRs on SI-GaAs.}%
\label{Figure4}%
\end{center}
\end{figure}

One consequence of embedding metallic resonators into nonlinear dielectrics
is the enhancement of the effective nonlinearity by 2
to 4 orders of magnitude (in comparison to the bare nonlinear material) because
of the in-gap field enhancement \cite{rose11b}. For our metamaterials, the
nonlinearity results in a change of the oscillator strength of
the LC resonance.  Because of dispersion, this leads to
changes in the effective permittivity and refractive index at
frequencies away from resonance. The effective real
permittivity ($\epsilon_{re}$) is shown as a function of frequency for various incident electric fields for
SRRs on n-type GaAs (Fig. \ref{Figure4}(a)) and SI-GaAs (Fig. \ref{Figure4}(c)). These plots
were obtained by performing parameter extraction (for a cubic unit
cell) using the full-wave simulations \cite{smith02,smith05,tao08} of Fig. \ref{Figure2}(c), (d) and
\ref{Figure3}(b) and subsequently mapping the in-gap conductivity to
the corresponding incident peak electric field required to achieve
that conductivity. It is clear that the field-dependence of $\epsilon_{re}$ is
different for the two cases. For the SRRs on n-type GaAs, the
off-resonance $\epsilon_{re}$ at 0.82 THz changes from positive to
negative owing to IVS in the doped thin film. Above 160 kV/cm, $\epsilon_{re}$ increases from
negative to positive values due to the onset of IMI (Fig. 4(b)). For SRRs on SI-GaAs, IMI
increases $\epsilon_{re}$ monotonically from negative to
positive values at 0.85 THz (Fig. 4(d)).

We have demonstrated that metamaterials enable strongly
enhanced nonlinearities at terahertz frequencies. The present
study illustrates strong THz-induced changes in the carrier mobility
and density, yielding conductivity changes up to ten orders of
magnitude and large changes in THz transmission. As our proof-of-principle results demonstrate,
it will be possible to create a host of nonlinear THz metamaterials which includes, as but
one example, nonlinear absorbers for saturable absorber or
optical limiting applications.

The authors acknowledge support from ONR grant no. N00014-09-1-1103, AFOSR grant no. FA9550-09-1-0708, and from
DTRA C$\&$B Technologies Directorate administered through a subcontract from ARL.
We would also like to thank Grace Metcalfe and Mike Wraback of ARL for providing the MBE grown n-type GaAs.

\end{document}